\documentclass{aa}  

\usepackage{graphicx}
\usepackage{txfonts}
\usepackage{hyperref}
\hypersetup{colorlinks=true,linkcolor=blue,citecolor=blue,urlcolor=blue}
\begin{document}

   \title{Fluorine Evolution in the Galactic Halo}


   \author{V. Grisoni
          \inst{1,2}
          \and
          F. Rizzuti \inst{3,4,1}
          \and
          G. Cescutti \inst{5,1,2,4}\fnmsep
          }

   \institute{INAF, Osservatorio Astronomico di Trieste, via G.B. Tiepolo 11, I-34131, Trieste, Italy\\
              \email{valeria.grisoni@inaf.it}
         \and
         IFPU, Institute for Fundamental Physics of the Universe, Via Beirut 2, 34151 Trieste, Italy
         \and
            Heidelberger Institut für Theoretische Studien, Schloss-Wolfsbrunnenweg 35, D-69118 Heidelberg, Germany
         \and
            INFN, Sezione di Trieste, via Valerio 2, I-34134 Trieste, Italy
        \and
        Dipartimento di Fisica, Sezione di Astronomia, Università di Trieste, Via G. B. Tiepolo 11, 34143 Trieste, Italy
             }

   \date{Received --; accepted --}

 
  \abstract
   {The chemical evolution of fluorine is still a matter of debate in Galactic archaeology, especially at low metallicities, where it is particularly challenging to obtain the corresponding chemical abundances from observations.}
   {We present here the first detailed theoretical study of the chemical evolution of fluorine at low metallicity by means of a stochastic chemical evolution model for the Galactic halo, in light of the most recent data for fluorine, which further pushed observations to lower metallicities down to [Fe/H]$\sim$-4 dex, more than a factor of 10 lower in metallicity than previous detections.}
   {We employ a state-of-the-art stochastic chemical evolution model to follow the evolution in the Galactic halo, which has been shown to reproduce well the main observables in this Galactic component and the abundance patterns of CNO and neutron-capture elements, and we implement nucleosynthesis prescriptions for fluorine, focusing on the chemical evolution of this element.}
   {By comparing recent observations with model predictions, we confirm the importance of rotating massive stars at low metallicities to explain both the [F/Fe] vs [Fe/H] and [F/O] vs [O/H] diagrams. In particular, we showed that we can reach high [F/Fe]$\sim$2 dex at [Fe/H]$\sim$-4 dex, in agreement with recent observations at the lowest metallicity.}
   {With a stochastic chemical evolution model for the Galactic halo, we confirm the importance of rotating massive stars as fluorine producers, as hinted by previous studies using chemical evolution models for the Galactic disc. We also expect an important production of F at high redshift, in agreement with recent detections of supersolar N by JWST. Further data for fluorine at low metallicities and also at high redshift would be needed to put further constraints on the chemical evolution of fluorine and be compared to our theoretical predictions.}


   \maketitle
%

\section{Introduction}

The origin and evolution of fluorine still represents a matter of debate in the field of Galactic Archaeology \citep[see e.g.][]{Ryde2020b}, especially at the early stages of
chemical enrichment in the Universe \citep{Ryde2021}.
Fluorine is very fragile and it has only one stable isotope, $^{19}$F. In literature, several stellar sites have been proposed for the production of fluorine, which can be summarized as follows: i) thermal pulses in asymptotic giant branch (AGB) stars \citep{Forestini1992,Jorissen1992,Straniero2006,Abia2011,Gallino2010,Karakas2010,Cristallo2014}, ii) Wolf-Rayet (WR) stars \citep{Meynet2000}, iii) rotating massive stars \citep{Limongi2018,Prantzos2018}, iv) the neutrino-process in core-collapse supernovae (CCSNe) \citep{Woosley1988,Kobayashi2011nu}), and v) novae \citep{Jose1998}.
\\In the literature, Galactic chemical evolution (GCE) models have been widely used to study the evolution of fluorine in the Milky Way and the role of the different fluorine producers \citep{Timmes1995,Meynet2000,Renda2004,Kobayashi2011base,Prantzos2018,Spitoni2018,Olive2019,Grisoni2020b,Womack2023}.
In the study of \cite{Timmes1995}, they showed that in principle the
$\nu$-process could provide a possible explanation for the origin
of fluorine, even if their yields of fluorine from core-collapse
supernovae including the $\nu$-process were not enough to reproduce the observational data. \cite{Meynet2000} suggested that WR stars can contribute significantly to the solar abundance of fluorine. Then, \cite{Renda2004} demonstrated quantitatively the impact of fluorine nucleosynthesis in Wolf-Rayet and AGB stars, and the importance of the inclusion of these latter two fluorine production sites. However, the contribution of WR stars to the chemical evolution of fluorine has been questioned in \cite{Palacios2005}. \cite{Kobayashi2011base} found that the contribution from AGB stars to the chemical evolution of fluorine can be seen only at [Fe/H]>-1.5 dex, and it is not enough to reproduce the observations at [Fe/H] $\sim$ 0 alone 
so
other fluorine sources should be needed. In particular, \cite{Kobayashi2011nu} showed that 
the $\nu$-process of core-collapse supernovae and the AGB stars combined together can give a significant contribution to the production of fluorine: the main impact of the $\nu$-process in the [F/O] vs. [O/H] plot is represented by the presence of a plateau, followed by the rapid increase due to AGB stars (see also \citealt{Olive2019}). Moreover, novae have also been included in chemical evolution models following the evolution of fluorine; in particular, \cite{Spitoni2018} showed that novae can contribute to better explain the observed secondary behavior of fluorine in the [F/O] vs. [O/H] diagram, even if their yields are very uncertain. Then, \cite{Prantzos2018} used the yields from rotating massive stars by \cite{Limongi2018} in a Galactic chemical evolution model, and they showed that this process can dominate the fluorine production up to solar metallicities. In particular, \cite{Grisoni2020b} focused on the chemical evolution of fluorine in the Galactic thick and thin discs, and they conclude that rotating massive stars are indeed important
producers of F below [Fe/H]=-0.5 dex,
though its contribution for [Fe/H]<-1 had yet to be confirmed by an extensive study in the Galactic halo; on the other hand, in order to reproduce the F abundance increase in the discs at late times,
a contribution from lower mass stars, such as single asymptotic giant branch stars
and/or novae, is required. Later on, also \cite{Womack2023} confirmed the importance of rotating massive stars in the chemical evolution of fluorine in the Milky Way, investigating the metallicity range -2$<$[Fe/H]$<$0.4. Still, a detailed GCE study of fluorine in the metal-poor regime is missing, as pointed out by \cite{Tsiatsiou2025} where they presented new results for the synthesis of F in massive stars with and without rotation, and stated that GCE models considering inhomogeneity of the interstellar medium (ISM) should be needed to follow fluorine evolution in the Galactic halo. In particular, a detailed stochastic chemical evolution model as presented in \cite{Rizzuti2025} can be applied to the study also of the chemical evolution of fluorine in the Galactic halo in the light of the most recent observational data, as we will show here.
\\From the point of view of observational data, in the past years, a great amount of work has been done regarding the chemical evolution of fluorine in the Galaxy \citep{Abia2011,Abia2015,Abia2019,RecioBlanco2012,DeLaverny2013a,DeLaverny2013b,Jonsson2014,Pilachowski2015,Guerco2019a,Guerco2019b,Ryde2020a,Nandakumar2023,Brady2024,Sesha2024a,Sesha2025b}.
There have been observations of fluorine in open and globular
clusters 
\citep{Cunha2003,Cunha2005,Smith2005,Yong2008,DeLaverny2013a,DeLaverny2013b,Nault2013,Maiorca2014,Guerco2019b,Sesha2024a,Sesha2025b}. There have been studies in different Galactic components, such as the Galactic thick and thin discs \citep{Guerco2019a,Ryde2020a,Brady2024}, the bulge \citep{Jonsson2014} and the  Galactic Nuclear Star Cluster  \citep{Guerco2022}.
Recently, \cite{Nandakumar2023} studied in detail the evolution of fluorine at high metallicities (namely, -0.9 $<$[Fe/H] $<$0.25 dex) and recommended a set of vibrational-rotational HF lines to be used in the abundance analysis.
On the other hand, the evolution of fluorine at low metallicity (e.g., [Fe/H] $<$ -1.5 dex)
poses a particular challenge because of a large contamination from telluric lines and blending of the HF lines with CO features \citep{Lucatello2011}. Despite those difficulties, there are some measurements of fluorine abundances at low metallicities ([Fe/H] $<$ -1.5 dex), such as in the work of \citealt{Lucatello2011,Li2013,Abia2015,Guzman2020,Guzman2025}.
In particular, in the very recent work of \cite{Guzman2025}, they presented F abundances and upper limits for 7 stars, pushing forward the limit of low metallicity in the literature for fluorine determination, reaching [Fe/H]=-3.87 dex, which is more than a factor of 10 lower in metallicity than previous detections. Finally, we note that fluorine has also been investigated at high redshift; in particular, \cite{Franco2021} were able to estimate the abundance of fluorine in a gravitationally lensed galaxy at a redshift of
z = 4.4, concluding that WR stars could be responsible for
the observed fluorine abundance enhancement; this can also give important information about the origin of
fluorine in the early Universe.
\\The aim of this paper is to model the evolution of fluorine in the Galactic halo by means of a stochastic chemical evolution model as recently presented by \cite{Rizzuti2025} in the light of the most recent data for fluorine at low metallicity (e.g. \citealt{Guzman2025}).

\section{Observational data}

\begin{table}
\centering
\caption{Observational data used in this work: in the first column we report the source, in the second column the abundances used.}
    \label{table}
    \begin{tabular}{l c}
        \hline
        Source & Abundances \\
        \hline \cite{Guzman2025} & F, Fe \\
        \cite{Abia2015} & F, Fe\\
        \cite{Lucatello2011} & F, Fe, O\\
        \hline
    \end{tabular}
\end{table}

In this work, we make use of recent data for fluorine in the metal-poor regimes, complemented by data from the literature. 
In particular, we consider the recent data from \cite{Guzman2025} where they presented F abundances and upper limits in 7 carbon-enhanced metal-poor (CEMP)  stars observed with the Immersion Grating Infrared Spectrometer (IGRINS) \citep{Yuk2010,Park2014,Mace2018}, at the Gemini-South telescope.
To complement this recent dataset, we also consider other results in the literature. In particular, we consider F abundances and upper limits from \cite{Lucatello2011}, where they presented results for a sample of 11 metal-poor red giants. Their chemical abundances were obtained from the analysis of spectra from the CRyogenic high-resolution cross-dispersed InfraRed Echelle Spectrograph (CRIRES) \citep{crires2004} at ESO Very Large Telescope (VLT). These stars are in the metallicity range -3.4<[Fe/H]<-1.3; two of them have measurements of fluorine, while for the others they present upper limits. We also consider Carina data from \cite{Abia2015}: even if these two are observations in Carina dSph galaxy, they can give hints about the evolution of fluorine at low metallicities. All the observational data used in this work are summarized in Table 1.

\section{Chemical evolution model}

\begin{figure*}
\centering
   \textbf{Rotating vs. non-rotating case}\par\medskip
 	\includegraphics[scale=0.56]{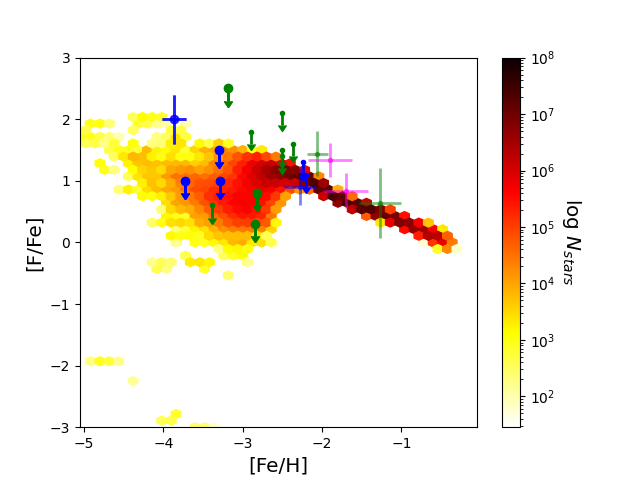}
    \includegraphics[scale=0.56]{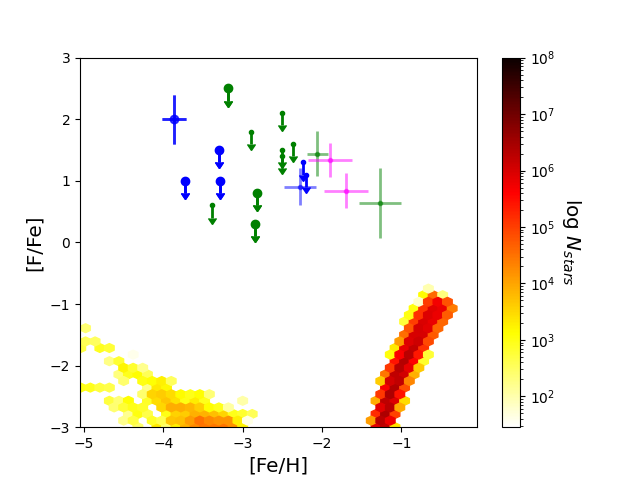}
    \caption{Observed and predicted [F/Fe] versus [Fe/H] for the Galactic halo for our reference chemical evolution model with variable rotational velocity as in \cite{Rizzuti2025} (left) and non-rotating massive stars (right). The data are taken from \cite{Guzman2025} (blue) and \cite{Lucatello2011} (green), both upper limits and measurements with corresponding error bars; smaller dots correspond to CEMP-s stars. We also show Carina data from \cite{Abia2015} (magenta).
    }
    \label{fluorine}
\end{figure*}

To follow the chemical evolution of fluorine in the Galactic halo, we adopt the stochastic chemical evolution code \texttt{GEMS} from \cite{Rizzuti2025}, where a detailed description of the model and choice of the parameters can be found.
\\In particular, in \cite{Rizzuti2025}, they presented a new version of the stochastic model first presented in \cite{Cescutti2010}, which was based on the original model of \cite{Cescutti2008} and on the homogeneous model of \cite{Chiappini2008}.
\texttt{GEMS} is specifically used to reproduce the chemical evolution of the Galactic halo, therefore it runs for 1 Gyr after the formation of the Galaxy. To reproduce inhomogeneities in the chemical composition, stochasticity is introduced by dividing the simulation domain into various subvolumes, each one having the typical volume covered by a type-II SN explosion, and all are considered as independent. In this work, thanks to the parallelized computing of the \texttt{GEMS} code that allows to increase the number of volumes, we consider a high number of volumes (here 1000, much higher than e.g. \citealt{Cescutti2008} where 100 volumes were considered). Further increasing the number of volumes would increase the coverage of rare events, but the high-probability areas predicted by the model would remain the same and thus the results would remain robust.
\\Only one infall episode is assumed for the Galactic halo, accreting gas of primordial composition with a Gaussian distribution (see e.g. \citealt{Chiappini2008});
$$\dot{G}(t)_{inf}=\frac{\Sigma_h A}{\tau \sqrt{2 \pi}} e^{-\frac{(t-t_0)^2}{2\tau^2}}$$
where t$_0$ is 100 Myr, $\tau$ is 50 Myr, and the normalization is given by $\Sigma_h$=80 M$_{\odot}$ pc$^{-2}$ and A is the surface of each cell, so that $\Sigma_h$A=3.2 x 10$^6$ M$_{\odot}$.
\\The initial mass function (IMF) is the \cite{Scalo1986}, as in \cite{Rizzuti2025}. In \cite{Rizzuti2025}, they also tested the IMF from \cite{Kroupa2001}, which is more
top-heavy compared to the one by \cite{Scalo1986}; they show the results in their Appendix (their Figs. C.1 and C.2), where the impact of the different IMF is not significant.
\\The star formation rate (SFR) in units of M$_{\odot}$ Gyr$^{-1}$, is defined as 
$$\psi(t)= \nu \Sigma_h A (\frac{G_{gas}(t)}{\Sigma_h A})^k$$
where $\nu$ is 1.4 Gyr$^{-1}$, k is 1.5, G$_{gas}$(t) is the amount of gas inside the volume in M$_{\odot}$.
\\Finally, it is considered a Galactic wind that takes gas out of each cell, with a rate proportional to the star formation:
$$\dot{G}(t)_{out}= \omega \psi(t)$$
with a constant $\omega$ set equal to 8 for all chemical species \citep{Chiappini2008}.
The evolution proceeds as follows. At each timestep, in each cell a certain mass of gas according to the SFR is converted into stars, randomly extracted with mass between 0.1-100 M$_{\odot}$. Thus, each cell has the same amount of stars in mass but with a different distribution. After stars are born, their evolution is followed until they die, considering the stellar lifetimes from \cite{Maeder1989}.
\\For further details on the model prescriptions, we address the interested reader to \cite{Rizzuti2025}, where an exhaustive description of the model can be found.

\begin{figure*}
\centering
   \textbf{Rotating vs. non-rotating case}\par\medskip
        \includegraphics[scale=0.56]{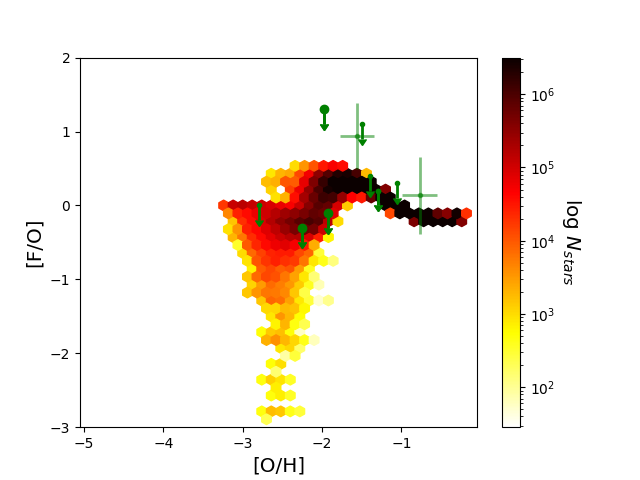}
    \includegraphics[scale=0.56]{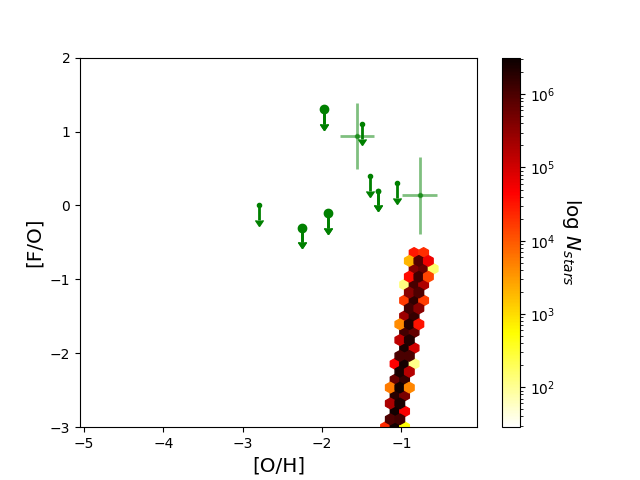}
    \caption{Same as Fig.~\ref{fluorine}, but for [F/O] vs. [O/H].
    }
    \label{fluorine_o}
\end{figure*}

\subsection{Nucleosynthesis prescriptions}
The nucleosynthesis prescriptions and the implementation of the
yields in the model are fundamental ingredients for chemical evolution models. In this work, we implement nucleosynthesis prescriptions following \cite{Rizzuti2025} with the inclusion of fluorine.
\\For rotating massive stars, we use the yields of \cite{Limongi2018} (their recommended set, set R), where they provided a grid of nine masses between 13-120 M$_{\odot}$, four metallicities [Fe/H]=0,-1,-2 and -3, and three rotations with initial equatorial velocity of 0 km s$^{-1}$ (non rotating), 150 km s$^{-1}$ and 300 km s$^{-1}$. This grid of models has been widely used in GCE studies (see for example \citealt{Prantzos2018,Prantzos2023,Rizzuti2021,Romano2020,Romano2021,Grisoni2021,Kobayashi2022,Rossi2024,Rizzuti2025}). There is still a large uncertainty in the explodability of massive stars, both observationally and theoretically. Still, the assumptions on the explosive mass range in our models have allowed in the past a good reproduction of the CNO and neutron-capture elements, as described in \cite{Rizzuti2025}. Therefore, we rely on these assumptions also for the reproduction of F in the present paper.
\\Following \cite{Rizzuti2025}, all massive stars are assumed to have the same rotation velocity depending on their metallicity:
$$\text{velocity([Fe/H])}=300 \text{ km s}^{-1} \text{for [Fe/H]} \le -3$$
$$\text{velocity([Fe/H])}=150 \text{ km s}^{-1} \text{for [Fe/H]} >-3 $$
This distribution is consistent with findings from \cite{Prantzos2018,Rizzuti2019}, with faster rotation in the early Galactic phases and slower in the subsequent ones (see also \citealt{Romano2020,Grisoni2021}). Here, we are interested only in metal-poor stars ([Fe/H]$<$-2) and therefore we do not decrease the rotation even further towards higher metallicity, as discussed in e.g. \cite{Rizzuti2021}.
\\To reproduce observations for [Fe/H]$<$-3 (the lowest metallicity in the grid of stellar models of \citealt{Limongi2018}), we consider the grid from \cite{Roberti2024}, where they also used the \texttt{FRANEC} code and provided 15 and 25 M$_{\odot}$ models at [Fe/H]=-4,-5 and zero metals (Z=0) for a wide range of rotational velocities. Further details on these sets of yields for fluorine are given in the Appendix.
\\Regarding the contribution from low- and intermediate-mass stars (LIMS), the yields have been assumed from stellar models obtained 
from the \texttt{FRUITY} database for non-rotating stars between 1.3 and 6 M$_{\odot}$ \citep{Cristallo2009,Cristallo2011,Cristallo2015}. However, their impact can be visible in the model results only from [Fe/H]$>$-2 dex.
Finally, for SNe Type Ia, the yields are from \cite{Iwamoto1999}, even if SNe Type Ia contribution to the fluorine budget is minimal.

\section{Results}

In Fig. \ref{fluorine}, we show the observed and predicted [F/Fe] vs. [Fe/H] for our reference model (left panel) and the non-rotating case (right panel).
\\We immediately note the very different predictions from the two cases (rotating vs non-rotating), and thus the importance of rotating massive stars as fluorine producers.
In particular, the model predictions for our reference model (left panel of Fig. \ref{fluorine}) are from the stochastic chemical evolution model of the Galactic halo of \cite{Rizzuti2025}, with the inclusion of fluorine nucleosynthesis as described in Section 3.1. The model of \cite{Rizzuti2025} has been shown to reproduce well the main observables in this Galactic component and the abundance patterns of CNO and neutron-
capture elements (see their Fig. 10 for their reference [O/Fe] vs. [Fe/H] diagram). 
\\Our model predictions for fluorine are compared to the data from \cite{Guzman2025} and \cite{Lucatello2011}. We also show data from \cite{Abia2015}; as previously mentioned, even if these two are observations in Carina dSph galaxy, they can give hints about the evolution of fluorine at low metallicities.
As already discussed in the Introduction, we note that there are many criticalities in the determination of fluorine abundances at low metallicity, and therefore many of the data available are only upper limits as indicated by the arrows. Even if upper limits are not as certain as absolute determinations, they can still inform us about the range of fluorine abundances that we should expect, and thus give us an idea of whether our predictions can match the observations. 
Also, we note that some stars in our sample are classified as CEMP-s (i.e. carbon-enhanced metal-poor stars enriched in s-process elements). These stars likely had their surface fluorine abundances altered by binary mass transfer from an AGB
companion changing the initial surface abundances (see \citealt{Beers2005,Lucatello2005,Bisterzo2010,Hansen2016}). Therefore, for such stars the predictions of our chemical evolution model can only provide a reference for the
average fluorine abundances at birth in CEMP-s stars, before
mass transfer took place.
Still, further data would definitely be needed to draw firm conclusions about fluorine evolution at lower metallicities. However, we highlight the novelty of our predictions in the light of the most recent data for fluorine available; such new data are pushing forward the limit in metallicity, as in the very recent work of \cite{Guzman2025}, presenting the lowest metallicity limit in fluorine available up to now. 
\\At the lowest metallicities, the model predictions for our reference model including rotating massive stars (left panel of Fig. \ref{fluorine}) show a tail reaching relatively high [F/Fe] values ([F/Fe]$\sim$2 dex at [Fe/H]$\sim$-4 dex, matching well the lowest metallicity F determination). The possibility to reach such high [F/Fe] is due to the fact that rotating massive stars produce fluorine and 
small amounts of iron
(see also [X/Fe] vs. [Fe/H] for CNO elements in \citealt{Rizzuti2025}, their Fig. 10). In this way, rotating massive stars can set high values for [F/Fe] at low metallicity, in good agreement with the data. We remind that rapidly rotating massive stars
can produce primary fluorine from $^{14}$N, through proton and alpha
captures in the presence of $^{13}$C \citep{Prantzos2018,Grisoni2020b}; $^{14}$N derives from reactions with
$^{12}$C, which comes from He burning in the massive star itself, and it is
then of primary origin. This chain of reactions
clearly happens also in the non-rotating case, but the available amount
of CNO nuclei is too small to contribute significantly to fluorine
production \citep{Limongi2018}.
\\We note that there are still no observations available for metallicity [Fe/H]$<$-4 dex; the lowest metallicity data is the very recent determination by \cite{Guzman2025} for CS 29498-0043, a CEMP-no star at [Fe/H]$=$ -3.87 with a F detection of [F/Fe] = +2.0 $\pm$ 0.4, the lowest metallicity star with observed F abundance to date, that we can well explain with our stochastic chemical evolution model.
\\For values -4$<$[Fe/H]$<$-2 dex, the model can reproduce the values for most stars, but we also note that we still have many upper limits.
\\Finally, for [Fe/H]$>$-2 dex, the contribution from LIMS becomes important but we do not extrapolate results from the model in this range, since our model is mainly focused on the Galactic halo. On the other hand, in  \cite{Grisoni2020b} they studied fluorine in the thick and thin discs by means of GCE models specifically tailored for investigating the Galactic discs \citep{Grisoni2017,Grisoni2018,Grisoni2019}, and they conclude that rotating massive stars are indeed important producers of F and they can set high values in [F/Fe] abundance below [Fe/H]=-0.5 dex (in agreement with \citealt{Prantzos2018,Womack2023}), though its contribution for [Fe/H]<-1 had to be confimed by an extensive study in the Galactic halo as done in the present paper.
On the other hand, in order to reproduce the F abundance increase in the discs at late times, a contribution from non-rotating massive stars, lower mass stars, single AGB stars and/or novae, should be required \citep{Spitoni2018,Grisoni2020b,Womack2023}.\\
In the right panel of Fig. \ref{fluorine}, we show the case with non-rotating massive stars and the difference is striking; we thus confirm the need for rotating massive stars to set high [F/Fe] values at low metallicities, not present in the non-rotating case.
\\
In Fig.~\ref{fluorine_o}, we then show the observed and predicted [F/O] vs [O/H], widely used in the literature to investigate the evolution of fluorine. In this way, we can remove the dependence 
on iron and use oxygen instead 
as a metallicity indicator. Also in this case, rotating massive stars can set a plateau in [F/O] at the lower metallicities 
as hinted by previous studies from the Galactic disc \citep{Grisoni2020b,Womack2023}. We can see that the model predictions for our reference model (left panel) can reproduce the general trends of the data (and upper values), at variance with the non-rotating case (right panel), where it is not possible to explain observations. Also in this case, thus we confirm the importance of rotating massive stars as fluorine producers.
Recently, in the paper of \cite{Tsiatsiou2025} (their Fig. 10), the authors present a [F/O]-[O/H] relation computed with the \texttt{CELIB} code \citep{Saitoh2017,Hirai2021}, showing a plateau at [O/H] $<$ –2 driven by rotating Pop III yields. Our results are consistent with their findings, stressing the importance of rotation to set a plateau at low [O/H].

\section{Summary and conclusions}

In this work, we have studied the chemical evolution of fluorine, which is still a matter of debate in Galactic archaeology, especially at low metallicity. To this aim, we present the first theoretical study of the chemical evolution of fluorine at low metallicity by means of a state-of-the-art stochastic chemical evolution model \citep{Rizzuti2025} in the light of the most recent data for fluorine at low metallicity \citep{Guzman2025}.
\\Our main results can be summarized as follows.
\begin{itemize}
\item With our stochastic chemical evolution model for the Galactic halo, we investigate fluorine evolution and the contribution from rapidly rotating massive
stars using the yields of \cite{Limongi2018} and \cite{Roberti2024}.
\item With such a set of nucleosynthesis prescriptions considering variable rotational velocity, it is possible to reach high [F/Fe] values $\sim$ 2 dex at [Fe/H]$\sim$-4 dex, in agreement with very recent observations from \cite{Guzman2025} where they first pushed observations to
lower metallicities down to [Fe/H]$\sim$-4 dex, at variance with the non-rotating case where it is not possible to reach such high values.
\item We confirm that
rotating massive stars can dominate the fluorine production at such low metallicities (in agreement with suggestions from other studies, e.g. \citealt{Prantzos2018,Grisoni2020b,Womack2023}).
\item We expect an important production of F also at high redshift (see e.g., \citealt{Franco2021}), in agreement with recent detections of supersolar N by JWST (e.g. \citealt{Cameron2023,Marques2024}), since also F is produced in the same fashion by both AGB and rotating massive stars, as we show here (see also \citealt{Kobayashi2024,Rizzuti2025b}).
\item On the other hand, to explain the secondary behaviour at higher metallicities ([Fe/H]$>$-2), other sites such as AGB stars and/or novae should be important, but we do not extrapolate results here since our model is tailored for the Galactic halo and not for the disc as done in detail in \cite{Grisoni2020b,Womack2023}.
\end{itemize}
In conclusion, we confirm that rotating massive stars can be important producers
of fluorine \citep{Prantzos2018,Grisoni2020b,Womack2023} and we show their impact on the chemical evolution of fluorine with a state-of-the-art chemical evolution model accounting for inhomogeneities in the ISM. Further data at low metallicities and at high redshift 
will be fundamental to further constrain the chemical evolution of fluorine.

\begin{acknowledgements}
We are grateful to the referee for all the useful comments and suggestions that improved our work.
VG acknowledges financial support from the INAF program “Giovani Astrofisiche ed Astrofisici di Eccellenza - IAF: INAF Astrophysics Fellowships in Italy" (Project: GalacticA, "Galactic Archaeology: reconstructing the history of the Galaxy") and from INAF Mini Grant 2023. FR is a fellow of the Alexander von Humboldt Foundation, and acknowledges support by the Klaus Tschira Foundation. FR also acknowledges I.N.A.F. for the 1.05.24.07.02 Mini grant 2024 “GALoMS - Galactic Archaeology for Low Mass Stars” (PI C.T. Nguyen). FR and GC acknowledge financial support under the National Recovery and Resilience Plan (NRRP), Mission 4, Component 2, Investment 1.1, Call for tender No. 104 published on 2.2.2022 by the Italian Ministry of University and Research (MUR), 570 funded by the European Union - NextGenerationEU - Project ‘Cosmic POT’ (PI: L. Magrini) Grant Assignment Decree No. 2022X4TM3H by the Italian Ministry of Ministry of University and Research (MUR).
G.C. thanks  I.N.A.F.
for the 1.05.23.01.09 Large Grant - Beyond metallicity: Exploiting the full POtential of CHemical elements (EPOCH) (ref. Laura Magrini)
and the European Union’s Horizon 2020 research and innovation programme for the grant agreement No 101008324 (ChETEC-INFRA).

\end{acknowledgements}


\bibliographystyle{aa}
\bibliography{aa_biblio}

\begin{appendix}

\section{Stellar yields}

\begin{figure*}
\centering
    \includegraphics[scale=0.5]{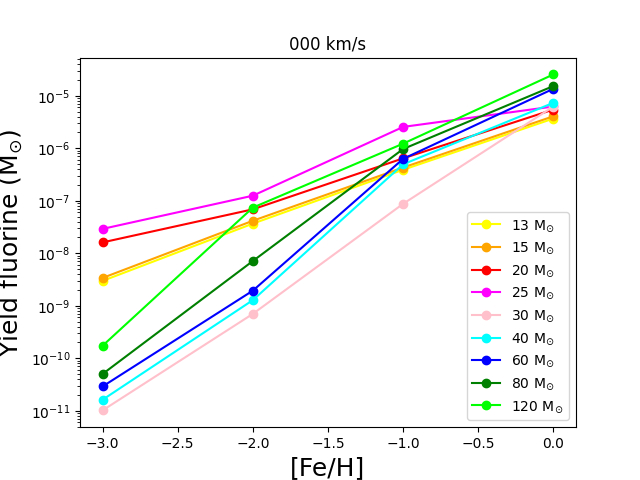}
    \includegraphics[scale=0.5]{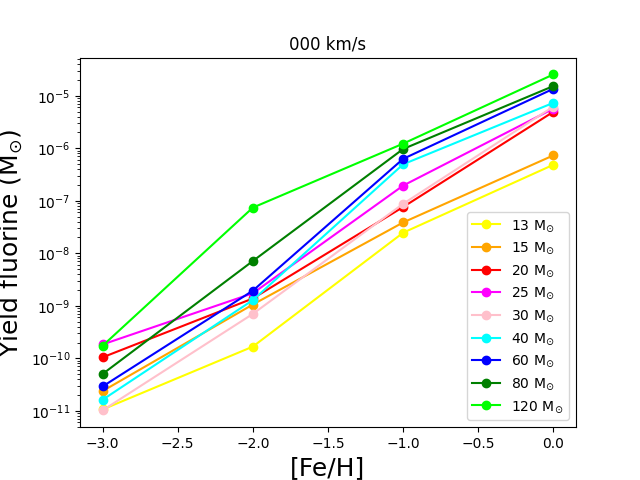}
    \includegraphics[scale=0.5]{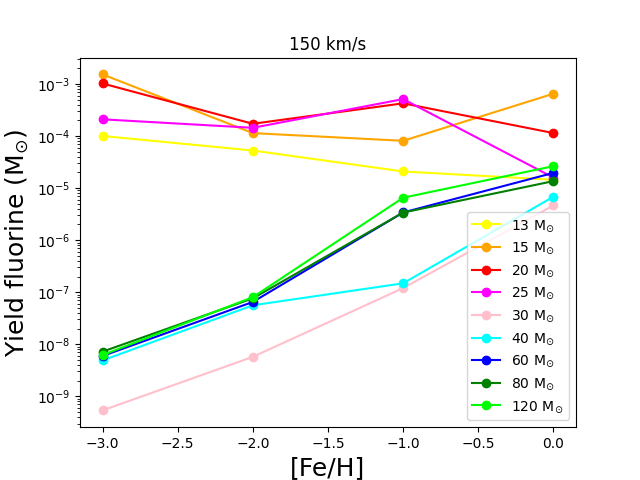}
    \includegraphics[scale=0.5]{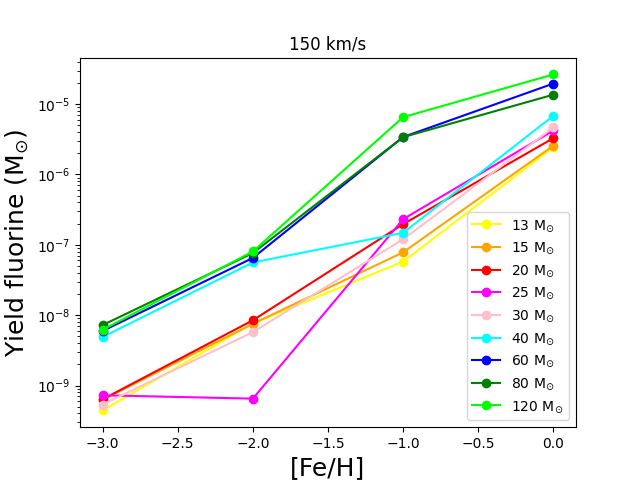}
    \includegraphics[scale=0.5]{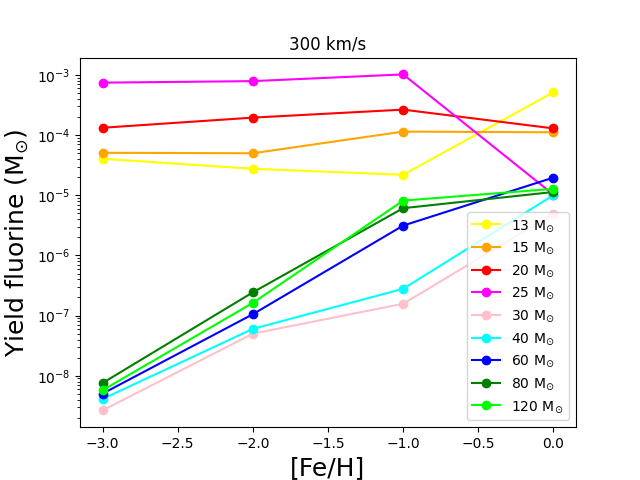}
    \includegraphics[scale=0.5]{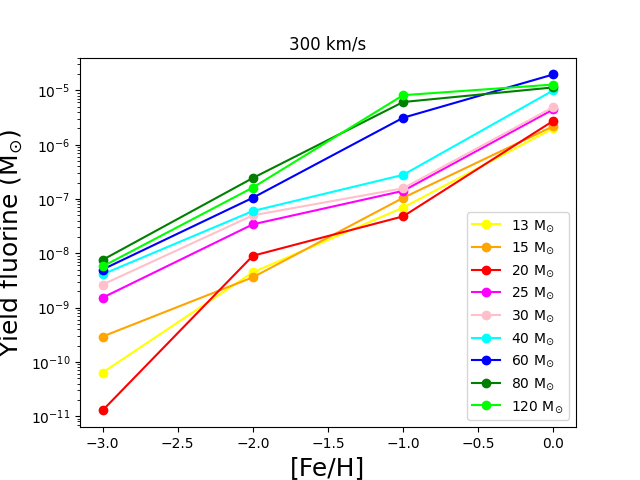}
    \caption{Fluorine total yields in the stellar models of \cite{Limongi2018} (left column) and only wind (right column), for three initial rotational velocities: 0 km/s (upper panels), 150 km/s (middle panels) and 300 km/s (lower panels).
    }
    \label{yields1}
\end{figure*}

\begin{figure*}
\centering
 	\includegraphics[scale=0.5]{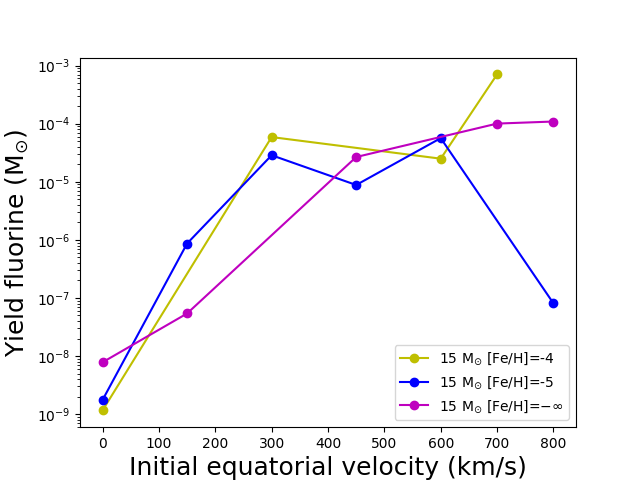}
    \includegraphics[scale=0.5]{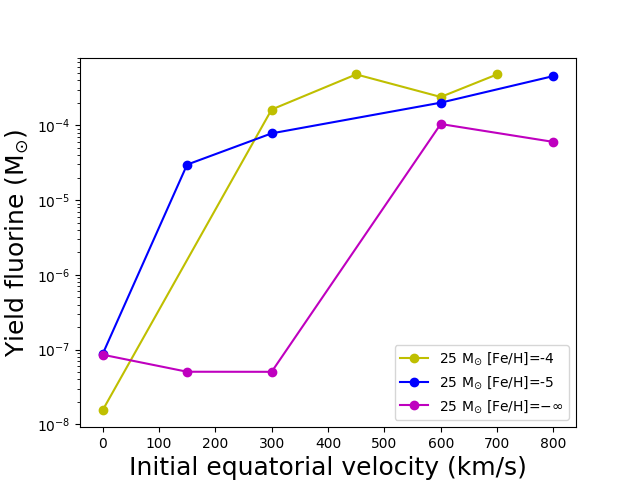}
    \caption{Fluorine yields in the stellar models of \cite{Roberti2024} for 15 (left) and 25 M$_{\odot}$ (right) stars, as a function of the star initial equatorial velocity in km/s. Different colours correspond to different metallicities: [Fe/H]=-4, -5 and $-\infty$ (zero metals).
    }
    \label{yields2}
\end{figure*}

Here, we discuss the stellar yields of fluorine from \cite{Limongi2018} and \cite{Roberti2024}, which have been included in our chemical evolution model for the Galactic halo. As mentioned in Section 3, \cite{Limongi2018} have presented a set of yields for different mass, metallicity, and rotational velocity.
\\In Fig. \ref{yields1}, we show the fluorine yields in the stellar models of \cite{Limongi2018} (left column) and only wind (right column), for three initial velocities: 0 km/s (upper panels), 150 km/s (middle panels) and 300 km/s (lower panels), set R and total yields. In particular, it is evident that fluorine is mostly produced by 13-25 M$_{\odot}$ at high rotational velocities (150-300 km/s). Those are the same stars that undergo supernova explosion in the set R. In Section 3, we also discussed that a new set of \texttt{FRANEC} models has been published by \cite{Roberti2024}, for 15 and 25 M$_{\odot}$ stars with metallicity [Fe/H]=-4,-5, and zero metals (Z=0) for various rotational velocities. Even if the mass range is small, the very low metallicity and large rotational velocity range make this grid a perfect tool for investigating the nucleosynthesis of the first stars (see \citealt{Rizzuti2025}).
In Fig. \ref{yields2}, we show the fluorine yields as predicted by the models of \cite{Roberti2024} as a function of the stellar initial equatorial velocity. From Fig.~\ref{yields2}, it is evident that fluorine generally increases with rotation, perfectly in agreement with our assumption of velocity function and \cite{Limongi2018}.

\end{appendix}

\end{document}